\documentstyle[amssymb,preprint,aps]{revtex}
 \tightenlines

\begin{document}
\title{Electronic Raman scattering on under-doped $Hg-1223$ high-Tc
superconductors: investigations on the symmetry of the order parameter}
\author{A. Sacuto, J. Cayssol , D. Colson$\dagger $ and P. Monod}
\address{Laboratoire de Physique de la Mati\`{e}re Condens\'{e}e, Ecole Normale\\
Sup\'{e}rieure, 24 rue Lhomond, 75231 Paris cedex 05 France.\\
$\dagger $Physique de l'Etat Condens\'{e}e, DRECAM/SPEC,CEA Saclay, 91191,\\
Gif sur Yvette, France.}
\date{Received \today }
\maketitle

\begin{abstract}
In order to obtain high quality, reliable electronic Raman spectra of a high-%
$Tc$ superconductor compound, comparable with existing model, we have
studied strongly under-doped $HgBa_{2}Ca_{2}Cu_{3}O_{8+\delta }$. This
choice was made such as to i) minimize oxygen disorder in the $Hg-$plane
generated \ by oxygen doping \ ii) avoid the need of phonon background
subtraction from the raw data iii)eliminate traces of parasitic phases
identified and monitored on the crystal surface. Under these experimental
conditions we are able to present the pure electronic Raman response
function in the $B_{2g}$, $B_{1g}$, $A_{1g}+B_{2g}$ and $A_{1g}+B_{1g}$
channels. The $B_{2g}$ spectrum exhibits a linear frequency dependence
(exponent $n=1.12$ $\pm $ $0.03)$ at low energy whereas the $B_{1g}$ one
shows a cubic-like dependence ( $n=3.02$ $\pm $ $0.1)$. The $B_{2g}$ and $%
B_{1g}$ spectra display two well defined maxima at $5.6$ $k_{B}Tc$ and $9$ $%
k_{B}Tc$ respectively. In mixed $A_{1g}$ channels an intense electronic peak
centered around $6.4$ $k_{B}Tc$ is observed. The low energy parts of the
spectra (below $350$ $cm^{-1}$) correspond to the electronic response
expected for a pure $d_{x^{2}-y^{2}}$ gap symmetry (and can be fitted up to
the gap energy for the $B_{1g}$ channel). However, in the upper parts (above 
$350$ $cm^{-1}$), the relative position of the $B_{1g}$ and $B_{2g}$ peaks
needs expanding the $B_{2g}$ Raman vertex to second order Fermi surface
harmonics to fit the data with the $d_{x^{2}-y^{2\text{ }}}$model. The
sharper and more intense $A_{1g}$ peak appears to challenge the Coulomb
screening efficiency present for this channel. As compared to previous data
on more optimally doped, less stoichiometric $Hg-1223$ compounds, this work
reconciles the electronic Raman spectra of under-doped $Hg-1223$ crystals
with the $d_{x^{2}-y^{2}}$model, provided that the oxygen doping is not too\
strong. This apparent extreme sensitivity of the electronic Raman spectra to
the low lying excitations induced by oxygen doping in the superconducting
state is emphasized here and remains an open question.
\end{abstract}


%\twocolumn

%\begin{multicols}{2}
%\narrowtext

\section{INTRODUCTION}

Most of the recent investigations of inelastic light scattering by free
carriers (electronic Raman scattering) on high-$Tc$ superconductors are
focused on the low energy electronic spectra (up to the gap energy) obtained
for different polarizations of the incident and scattered light. These
experiments are carried out in order to study the angular dependence of the
superconducting order parameter. Indeed, theoretical calculations initiated
by Deveraux et al.\cite{1}predicted specific power laws of the electronic
Raman intensity at low energy depending on the superconducting gap symmetry
and the light polarization. More precisely, for a $d_{x^{2}-y^{2}}$ gap (or
a $\left| \ d_{x^{2}-y^{2}}\right| $ gap since Raman scattering is not phase
sensitive), the $A_{1g}$ and $B_{2g}$ symmetries involved (deduced
respectively from \ parallel and crossed polarizations along Cu-O direction)
should display a linear frequency dependence whereas the $B_{1g}$ symmetry
(obtained from crossed polarisations at $45$ degrees from the Cu-O
direction) should exhibit a cubic frequency dependence below the gap energy.

In order to carry out with reasonable confidence such an investigation a
number of experimental difficulties must be overcome. The main reason for
the occurence of these difficulties stems out of the intrinsic weakness of
the electronic Raman intensity which tends to be easily hindered by other
contributions to the spectra. These originate from i) phonons scattering ,
ii) parasitic phases\ (luminescence or Raman) and iii) the effect of
disorder.

The aim of this paper is to present clean and reliable electronic Raman
measurements on high$-Tc$ superconductors. This implies to find the
appropriate high$-Tc$ superconductor which allow us to eliminate the extra
scattering contribution mentioned just above. We briefly discuss below each
of these various contributions together with their possible cure.

The first major experimental difficulty is that in most compounds e.g. $%
La_{2-x}Sr_{x}CuO_{4},$ $YBa_{2}Cu_{3}O_{7-\delta }$ $(Y-123)$, $%
Bi_{2}Sr_{2}CaCu_{2}O_{8+\delta }$ $(Bi-2212)$ and $Tl_{2}Ba_{2}CuO_{6+%
\delta }$ $(Tl-2201)$, an intense phonon structure is observed together with
the weaker continuum of the electronic excitations.\cite{2,3,4,5} In order
to study the pure electronic Raman response function the phonon peaks have
to be subtracted from the spectra. Vibrational structures are experimentally
defined from the N state spectra and removed from the S state one. Without
considering other contribution, this procedure cannot take into account the
phonon temperature dependence (changes on the lineshape and intensity). The
intrinsic weakness of the electronic Raman scattering, resulting in a
moderate signal to noise ratio, makes this procedure awkward. Despite these
difficulties, qualitative agreement has been found with the $d_{x^{2}-y^{2}}$%
model \cite{2,3,4,5} although the exponent for the power law of the $B_{1g}$
channel is often exceedingly dependent on the phonon subtraction protocole.
Raw data (before subtraction of phonons) frequently exhibit an ambiguous
exponent of the power law of the $B_{1g}$ channel. An attempt to avoid this
problem \cite{6} consisted to work on $Tl-2201$ superconductor, staying out
of resonance of the phononic scattering by the use of the red excitation
line. Qualitative agreement with the $d_{x^{2}-y^{2}}$model for the
stoichiometric compound was obtained from raw data (without subtraction of
phonon) in $B_{1g}$ channel.

In our previous work, in contrast with the above mentioned compounds, we
observed that the\ $HgBa_{2}Ca_{2}Cu_{3}O_{8+\delta }$ compounds exhibit
much lower intensity phonon lines (about one order of magnitude) for
polarization lying into the $CuO_{2}$ planes.\cite{7,8}. The decrease of the
Raman activity of the $B_{1g}$ phonon can be understood by symmetry
considerations with respect to the $CuO_{2}$ layer. Indeed it has been shown 
\cite{9} that the $B_{1g}$ Raman phonon activity increases with the crystal
electric field induced by the cationic planes above and below the $CuO_{2}$
plane. In the $Hg-1223$ mecrcurate the charge equivalence of the two cations
($Ba^{2+}$ and $Ca^{2+}$) \ minimizes the local electric field and thus
reduces the Raman phonon activity. Conversely in the $Y-123$ compound where
two differently charged cations ($Y^{3+}$, $Ba^{2+}$) are present above and
below the $CuO_{2}$ plane, the Raman phonon activity is enhanced.

This favourable situation allowed us to carry out analysis of the Raman
electronic spectra in the normal and superconducting states without dealing
with the delicate handling of ''phonon subtraction''. Another advantage to
work with mercurates is that these compounds belong to the highest $Tc$
cuprate family which provides a larger spectral range below the
superconducting gap frequency where the low-energy electronic spectrum can
be analyzed.

The second major difficulty is due to the presence of parasitic phases on
the surface of crystal mercurates. We have measured severals $Hg-1223$
single crystals from various batches corresponding to different oxygen
doping. Significant intrinsic variation of the shape of the low energy
spectrum (and thus changes of the power law) have been observed. These
changes occur not only from batch to batch but, surprisingly, may also
depend on the impact point of the laser beam onto the surface of the same
crystal. \cite{9a} As will be shown below, electronic microscopy analyses
revealed traces of $BaCuO_{2}$, $BaO_{2}$ and $HgO$ which appear on as grown
single crystal surfaces after exposure to air (for a few days). We
discovered by a specific study on each of these phases (in bulk form) that
their electronic background is able to modify significantly the spectra even
for minute quantities.

The third difficulty originates from oxygen doping. All studies carried out
on under-doped, optimally doped and over-doped high-$Tc$ superconductors
revealed that the low energy electronic spectra is very sensitive to the
oxygen doping.\cite{6,10,10a}. In contrast with the stoichiometric $Tl-2201$
compound, optimal doping of the $Hg-1223$ compound is reached for
over-stoichiometry of oxygen. However, excess of oxygen introduces disorder
into the $Hg-$plane. The effect of this disorder (as we will see) is
qualitatively to alter the electronic Raman spectrum at low energy, namely
introducing a linear-like frequency dependence visible in the $B_{1g}$
channel. In particular, our previous work\cite{10aa} on electronic Raman
spectra on nearly stoichiometric $Hg-1223$ crystals displayed a frequency
dependence of the $B_{1g}$channel in clear contradiction with the $%
d_{x^{2}-y^{2}}$ \ symmetry.

As a consequence, in order to avoid the phonon problem, to eliminate traces
of parasitic phases and reduce the oxygen disorder coming from $\delta $
over-stoichiometry, we have chosen to study strongly under-doped (UD) $%
Hg-1223$ crystals (nearly stoichiometric), cleaning the surface just before
Raman measurements.

Using these experimental conditions, we report pure electronic Raman spectra
in the N and S states of the strongly UD $Hg-1223$ single crystals. Our main
results are summarized as follows: in the S state, the $B_{2g}$ electronic
spectrum exhibits a linear frequency dependence at low energy up to $350$ $%
cm^{-1}$ whereas the $B_{1g}$ one shows a cubic frequency dependence up to $%
700$ $cm^{-1}$\cite{10b}. The Raman spectra of the $Hg-1223$ in the $B_{2g}$
and $B_{1g}$ channels show for the first time a well defined maxima at $5.6$ 
$k_{B}T_{c}$ and $9$ $k_{B}T_{c}$ respectively. Mixed $A_{1g}+B_{2g}$ and $%
A_{1g}+B_{1g}$ symmetries exhibit a strong maximum close to $6.4$ $%
k_{B}T_{c} $.

Below $300$ $cm^{-1}$ the electronic spectra of the strongly UD $Hg-1223$
for all the channels (and up to the gap energy for the $B_{1g}$) exhibit the
electronic scattering expected for a pure $d_{x^{2}-y^{2}}$ \ gap. The
quality of the experimental data offers a compelling evidence for the $d-$%
wave model. However to reconcile the $B_{1g}$ and $B_{2g}$ peak positions,
the $B_{2g}$ Raman vertices have to be expanded to the next higher order of
Fermi surface harmonics \cite{10c}. The strong maximum in mixed $%
A_{1g}+B_{2g}$ and $A_{1g}+B_{1g}$ channels cannot be simply explained by
Coulomb screening. These two last points raise new questions about
significance (within the $d_{x^{2}-y^{2}}$ framework) of the electronic
maxima detected on the Raman spectra and the effective Coulomb screening of
the $A_{1g}$ channel.

\bigskip

\section{\protect\bigskip Crystal and surface characterisation}

The crystallographic structure of $Hg-1223$ is tetragonal and belongs to the 
$^{1}D_{4h}$ space group \cite{10d} which allows us a direct comparison with
theoretical calculations based on tetragonal symmetries. $%
HgBa_{2}Ca_{2}Cu_{3}O_{8+\delta }$ single crystals were grown by a single
step synthesis as previously described \cite{11,12a} resulting in strongly
under-doped system with only few stacking defects. Oxygen doping (when
needed) was achieved by a separate oxygen annealing yielding slightly
underdoped system. They are parallelepipeds with typical 0.7 x 0.7 $mm^{2}$
square cross section and thickness 0.3 mm.

The [100] crystallographic direction lies at $45$ degrees of the edge of the
square and the [001] direction is normal to the surface. This geometry helps
us to align the crystallographic directions with respect to the light
polarization \cite{12aa}. DC magnetization measurements on strongly and
slightly UD $Hg-1223$ crystals under field cooling (FC) and zero field
cooling (ZFC) are displayed in Fig.1. A sharp transition occurs at $%
T_{c}\sim $ $121$ $K$ , ($\Delta Tc\thicksim 5$ $K)$ for the stongly UD\
mercurate whereas the slightly UD $Hg-1223$ exhibits a larger transition\
width ($\Delta Tc\thicksim 10$ $K)$ with onset of diamagnetism around $133$ $%
K$ and a main transition close to $126$ $K$. This most likely manifests a
variation of the oxygen content inside the slightly UD crystals. We estimate
the oxygen doping from the $Tc-\delta $ diagram recently established \cite
{12b}. We find $\delta =0.12$ for $Tc=121$ $K$ whereas $Tc=126$ $K$
corresponds to $\delta =0.16$\cite{12bb}. Up to now, the smallest value of $%
\ \delta $ experimentally obtained is close to $0.1$.\cite{12b,12c} Optimal
doping is reached for{\it \ \ }$\delta =0.19$. Since over stoichiometry ($%
\delta $) refers to an excess of oxygen atoms randomly arranged within the $%
Hg-$ planes, optimal doping implies oxygen disorder in the $Hg-$ plane. In
order to reduce the oxygen disorder inside the $Hg-1223$ structure it is
therefore more appropriate to study UD $Hg-1223$ crystals. Moreover the UD $%
Hg-1223$ system is closer to a four-fold symmetry than an optimally doped
one and is thus more appropriate to symmetry considerations. Resistivity
values at $T_{c}$ is typically $\varrho (T_{c})\sim (30-60)\mu \Omega .cm$ 
\cite{12d} showing the presence of very little (if any) impurities or
defects. A\ particular attention has been paid on the surface quality as
already mentioned above. The mirror surface of the freshly as-grown single
crystal observed with optical microscope reveal $0.1$ $mm^{2\text{ }}$%
defect-free domains. However, after a few days of exposure to atmosphere new
phases appear due to local decomposition. These new phases undetectable by
X-ray diffraction measurements, cluster together ($10$ $\mu m$) and are
identified as $BaCuO_{2}$, $BaO_{2}$, $HgO$ and $Ca_{0.8}CuO_{2}$ by a local
electron probe analysis. In order to eliminate these phases we have polished
the UD $Hg-1223$ single crystal surface just before Raman measurements.$\ $%
The single crystal is fixed by beeswax and polished by hand using diamonds
paste of $\ 0.1\mu m$ \ and the adequate polishing felt disc. The crystal is
then cleaned with aceton for spectroscopy (containing less than $0.1\%$ of $%
H_{2}O$). We thus regenerate the original high quality defect-free surface
of the crystal without local decomposition. Sometimes one or two inclusions
of $Ca_{0.8}CuO_{2}$ $(40\mu m)$ are observed in the bulk and disapear upon
further polishing. Scanning tunneling measurements on the polished surface
reveals that the polishing streaks are less than $0.01\mu m$ deep. An
important optical test is that polished crystal surface under crossed
polarization remains dark. This shows that the mechanical polishing does not
alter the polarization of the incident light.

\section{Experimental details}

Raman measurements were performed with a double monochromator (U-1000
Jobin-Yvon) using a single channel detection (EMI 9863B photomultiplier) and
the $Ar^{+}$ laser 514.52 nm line. The spectral resolution was set at $2$ $%
cm^{-1}.$ The crystals immediately after polishing were mounted (in the
vacuum) on the cold finger of a liquid helium flow cryostat. The temperature
was controlled by a CLTS resistance located on the backside of the cold
finger. The incident laser spot is less than 100 $\mu m$ in diameter and the
intensity onto the crystal surface was kept below $25$ $W.cm^{-2}$ in order
to avoid heating of the crystal. The incidence angle was $60$ degrees. The
polarization is denoted in the usual way: x [100] (a axis), y [010],
x'[110], y'[1$\bar{1}$0]. In order to compare our experimental data with the
theoretical calculations,\cite{1} the pure $B_{2g}$ (xy) and $B_{1g}$ (x'y')
symmetries are needed. This requires that the incident electric field have
to lie within the xy plane. The incoming light being not normal to the
surface, the crystal mount has to be rotated by $45$ degrees \ between the $%
B_{1g}$ and $B_{2g}$ spectra. The $A_{1g}+B_{1g}$ and $A_{1g}+B_{2g}$
symmetries are obtained from the (xx) and (x'x') polarizations respectively.
In order to probe the same area for all symmetries, the impact of the laser
beam onto the crystal surface was precisely located before turning the
crystal. Only large homogenous surfaces have been selected. The least
diffusive spot was selected in order to \ minimize the amount of spurious
elastic scattering. Each Raman spectrum has been obtained after a sum of 40
scans. Each scan has spanned between $25$ and $1000$ $cm^{-1}$ with an
increment of $2$ $cm^{-1}$ and an integration time of $3$ $s$.

\bigskip

\section{Experimental Results}

In view of the presence of the extra phases, we have performed systematic
investigations of the hitherto non reported Raman spectra of the parasitic
phases related to mercury compounds, i.e : $BaO_{2}$, $BaCuO_{2}$, $%
Ba_{2}CuO_{3}$, $HgO$, $HgBaO_{2}$, $CaCO_{3}$. Measurements were carried
out in the bulk form (powder) in the full range of interest (up to $1000$ $%
cm^{-1}$) and in the same experimental conditions as the one of the $B_{1g}$
and $B_{2g}$ spectra in the superconducting state (excitation line: 514.52
nm, light under crossed polarization , T$=13$ $K$). The results will be
presented separetly\cite{13} but can be summarized as follows:\ All these
phases exhibit a nearly flat background together with caracteristic phonon
peaks \cite{13a} except the $BaO_{2}$ phase which shows a linear frequency
dependent background extending up to $1000$ $cm^{-1}$. This last phase can
significantly alter the analysis of Raman spectra as will be shown.

Imaginary part of the Raman response function in the S state and the N state
of the strongly UD $Hg-1223$ compound for the $B_{2g}$, $B_{1g}$, $%
A_{1g}+B_{2g}$ and $A_{1g}+B_{1g}$ channels are displayed \ in Fig.2. The
imaginary part of the susceptibilities in the superconducting and the normal
states are obtained from Raman measurements at $T=13$ $K$ and $T=295$ $K$
respectively. We present in Fig.2 the most representative spectra of a
serial of three sets of measurements perfomed on a strongly UD $Hg-1223$
crystal. In the following order, we have subtracted the dark current of the
photomultiplier \cite{13aa}, the Rayleigh scattering when present, \cite{13b}
corrected the spectra for the response of the spectrometer and the detector\
and divided the spectra by the Bose-Einstein factor $n(\omega ,T)=[exp(\hbar
\omega /k_{B}T)-1]^{-1}$. For the $B_{1g}$ and $B_{2g}$ channels obtained
from crossed polarizations the response was scaled up by a factor 1.2 with
respect to the $A_{1g}$ mixed channels in order to take into account the
absorption of the half wavelength plate inserted between the analyzer and
the entrance slit of the spectrometer. The insertion of the half wavelength
plate allows us to restore the same polarization inside the spectrometer for
all the channels and thus benefit from the best configuration with respect
to the grating factor efficiency.

In the S state, we clearly observe a linear frequency dependence at low
energy (exponent $n=1.12$ $\pm $ $0.03$ from Fig.2) with a well defined
electronic maximum near $450$ $cm^{-1}$ for the $B_{2g}$ channel whereas the 
$B_{1g}$ channel exibit no measurable electronic scattering below $200$ $%
cm^{-1}$in the limit of the photomultiplier detection and a non linear
frequency dependence (exponent $n=3.02$ $\pm $ $0.1$ from Fig.2) until we
reach a maximum around $800$ $cm^{-1}$ \cite{10b}. We observe in Fig.2, for
both $B_{1g}$ and $B_{2g}$ channels, that the intensity of the electronic
continuum above $800$ $cm^{-1}$in the N state is two times smaller than for
the S state. Temperature dependence of the optical absorption $2\kappa
(T)\omega _{L}/c$ ($\omega _{L}\approx 19000$ $cm^{-1})$ between $13$ $K$\
and $295$ $K$ cannot explain such a large difference since the infra-red
reflectance spectra is temperature independent above $1000$ $cm^{-1}$\cite
{13c} .

In the $A_{1g}$ mixed symmetries, the electronic continuum shows a linear
increase with respect to $\omega $ with a maximun centered around $550$ $%
cm^{-1}$. The slope of the electronic scattering is twice larger in the $%
A_{1g}+B_{2g}$ spectrum than in the $A_{1g}+B_{1g}$ one. This can be
attributed to the $B_{2g}$\ linear contribution which increases the slope at
low energy and broadens the low energy side of the $A_{1g}+B_{2g}$ maximum.
As a consequence the $A_{1g}+B_{1g}$ peak is narrower than the $%
A_{1g}+B_{2g} $ one. A special remark can be made concerning the variation
of the intensity of the phonon peaks at $540$ and $590$ $cm^{-1}$ in the S
state in comparison with the N state. The intensities of these two phonon
peaks are reduced by a factor of 3 when temperature raises from $T=13$ $K$
to $150$ $K$ and decrease again by a factor of 1.2 between $150$ $K$ and $%
295 $ $K.$ The strong increase of the phonon intensity is clearly a
superconductivity related effect since the phonon exhibit two regimes of
intensity variations above and below $Tc$. The Raman phonon activities
become similar to those obtained when electric fields are perpendicular to
the xy plane i.e. in zz polarization. The intensity enhancement of the
phonons in the superconducting state has also been observed in other
cuprates and a modified Raman scattering mechanism induced by a
redistribution of the excitated states in the superconducting state has been
proposed.\cite{14}

\bigskip

\section{\protect\bigskip Comparison with previous works}

In our previous work on slightly UD $Hg-1223$ mercurates, contrary to the
present work on strongly UD $Hg-1223$, we reported a linear frequency
dependence at low energy for the $B_{1g}$ spectrum (as well as for the $%
B_{2g}$) and one extra shoulder near $800$ $cm^{-1}$\ for\ the $A_{1g}$
spectrum beside the $550$ $cm^{-1}$ peak. \cite{10aa}\ These observations,
in contradiction with the $d_{x^{2}-y^{2}}$ model, have been interpreted as
the signature of an order parameter of lower symmetry. In view of the
specific study on the parasitic phases mentioned above, we can now assign
the presence of this shoulder to the presence of the $BaCuO_{2}$ phase. This
is also supported by the observation on the $A_{1g}$ spectrum of the intense 
$640$ $cm^{-1}$phonon peak which is another signature of the $BaCuO_{2}$
phase. In the same way, we suspect that the multicomponents features
observed by Zhou et al. \cite{14b} on the Raman spectra of $Hg-1223$ single
crystals originate from vibrational modes of $HgO$ and $BaCuO_{2}$ traces
present on the crystal surface. Indeed the $340$ $cm^{-1}$ and $640$ $%
cm^{-1} $ peaks are characterisitc of $HgO$ and $BaCuO_{2}$ spectra and
correspond to the oxygen motions of the $HgO$ and $BaCuO_{2}$ phases.
Concerning the linear background previously reported \cite{10aa} two
explanations can be proposed a priori : i) scattering related to parasitic
phases on the crystal surface, ii)\ scattering induced by oxygen doping.
Among different phases we quoted above, only the $BaO_{2}$ phase which
provides frequency linear background and is able to compete even with minute
quantities, with the superconducting \ signal \cite{13}.\ Quantitative
estimate \cite{14c} show that less than $1\%$ of $BaO_{2}$ is sufficient to
produce scattering of the order of magnitude of the electronic Raman coming
from the S state. It is therefore mandatory to thoroughtly eliminate these
phases before any optical investigation. However an additionnal $BaO_{2}$
background which extends beyond to $1000$ $cm^{-1}$ makes it difficult to
interpret the decrease of the continuum above $800$ $cm^{-1}$observed on
slightly UD mercurates. Moreover we observed that the temperature dependence
of the electronic Raman changed at $Tc$ transition\cite{10aa}.For these
reasons we mainly attribute the linear frequency dependence at low energy in 
$B_{1g}$ channel to the intrinsic effect of oxygen over-stoichiometry rather
than only to the extrinsic $BaO_{2}$ phase. Many Raman experiments have been
performed on high$-Tc$ superconductors as a function of oxygen doping which
reveal changes in the low energy electronic continuum. \cite{6,10,10a}
Optimaly doped (stoichiometric) $Tl-2201$ compounds exhibit a cubic like
dependence at low energy whereas the overdoped (over-stoichiometric) $%
Tl-2201 $ compounds show a linear dependence at low energy.\cite{6} Impurity
scattering was proposed to explain such a behaviour\cite{14d}. In our
previous data of slightly UD $Hg-1223$ crystal the linear background extends
up to $600$ $cm^{-1}$ which means a scattering impurity rate $\Gamma $ of
the order of the gap $\Delta _{0}$ itself. Such a high scattering rate
should strongly reduce the critical temperature. This is in contradiction
with $Tc=126$ $K$ and resisitivity value ($\rho (Tc)=(30-60)\mu \Omega cm$.
We propose that the linear background is the (direct or indirect) effect of
localized states generated by oxygen disorder in the $Hg-$plane. As
mentioned in section II, the oxygen disorder is expected to be more
important in nearly optimally doped or slightly UD system than the strongly
UD one. This is supported by infra-red reflectance measurements \cite{13c}.
Indeed a strong scattering rate at low energy ($T=15$ $K$) is observed in
nearly optimally doped system as well as an intense peak (near $184$ $%
cm^{-1})$ on the real part of the optical conductivity. This peak is often
seen in high-$Tc$ superconductors in presence of oxygen disorder\cite
{13c,14e}. Conversely, in strongly UD system there is no residual scattering
at low frequency ($T=15$ $K$) and the $184$ $cm^{-1}$peak is narrower and
weaker.

\section{Model for electronic Raman scattering}

We compare our experimental data with the $d_{x^{2}-y^{2}}$ model. In the
limit where the superconducting correlation length is much smaller than the
optical penetration depth, the imaginary part of the response function for $%
T\longrightarrow 0$ \ including the screening due to Coulomb interaction is
\ given by : \cite{16} 
\[
{\chi }^{"}(\omega )=Re\left[ {{\chi }_{\gamma \gamma }(\omega )-{\frac{{%
\chi }_{\gamma 1}(\omega ){\chi }_{1\gamma }(\omega )}{{\chi }_{11}(\omega )}%
}}\right] \text{ \ \ \ \ \ \ \ \ \ \ \ \ \ \ \ \ \ \ \ \ \ \ \ \ \ \ (1)} 
\]
where ${\chi }_{\gamma \delta }$ is defined by : 
\[
{\chi }_{\gamma \delta }(\omega )={\frac{2\pi {N}_{F}}{\omega }}\left\langle 
{{\frac{{\gamma }_{{\bf k}}{\rm \delta }_{{\bf k}}^{\star }{\rm \Delta }_{%
{\bf k}}^{{\rm 2}}}{{({\omega }^{2}-4{\Delta }_{{\bf k}}^{{\rm 2}})}^{1/2}}}}%
\right\rangle \text{ \ \ \ \ \ \ \ \ \ \ \ \ \ \ \ \ \ \ \ \ \ \ \ \ \ \ \ \
\ \ (2)} 
\]
Here ${\gamma }_{{\bf k}}$ and ${\delta }_{{\bf k}}$ are general Raman
vertices already described \cite{10aa}. $N_{F}$ is the density of states for
both spin orientations at the Fermi level, and the brackets indicate an
average over the Fermi surface. $\Delta _{{\bf k}}$ stands for the
superconducting, ${\bf k}$-dependent gap. Eq.(2) vanishes in the limit of
large frequency where we should recover the electronic scattering of the
normal state. Therefore the normal state scattering contribution is not
taken into account in the Raman response function derived in this work.\ 

In the $d_{x^{2}-y^{2}}$ model, as computed by Devereaux et al.\cite{1,2},
we approximate the Fermi surface as being a cylinder and take: $\Delta _{%
{\bf k}}$ $=\Delta _{0}Cos(2\theta )$ for the gap and for the vertices:$\ \ $

$\gamma _{k}^{B_{1g}}=\gamma _{B_{1g}}Cos(2\theta ),$ $\ \gamma
_{k}^{B_{2g}}=$ $\gamma _{B_{2g}}Sin(2\theta ),$ $\gamma
_{k}^{A_{1g}}=\gamma _{0}+\gamma _{A_{1g}}Cos(4\theta )$ \ \ \ \ \ \ \ \ \ \
\ \ \ \ \ \ \ \ \ \ \ \ \ \ \ \ \ \ \ (3)

In $B_{1g}$ and $B_{2g}$ symmetries, $\chi _{\gamma 1}$ vanishes which
implies that the last term in Eq.(1) drops out . In other words, the $B_{1g}$
and $B_{2g}$ symmetry are not screened. This is not the case for the $A_{1g}$
symmetry, and screening has to be properly taken into account. Nevertheless
one sees easily from this equation that any {\bf k}-independent contribution
to the Raman vertex drops out of the result.

In a pure $d_{x^{2}-y^{2}}$ model, the $B_{1g}$ channel is \ not sensitive
to the nodes because the Raman vertex $\gamma _{{\bf k}}$ is zero by
symmetry for $k_{x}=\pm k_{y}$ in contrast, the $k_{x}=0$ and $k_{y}=0$
regions do contribute, giving weight to the gap $\Delta _{0}$ in these
directions. As a consequence the $B_{1g}$ spectrum displays a weak
scattering at low energy and $\ $the calculated spectrum shows a $\omega
^{3} $ frequency dependence with a maximum at $2{\Delta }_{0}$. Conversely,
in the $B_{2g}$ channel, the Raman vertex $\gamma _{{\bf k}}$ is zero by
symmetry for $k_{x}=0$ or $k_{y}=0$ or and non zero elsewhere, hence
provides weight in the $k_{x}=\pm k_{y}$ directions where the nodes are
present. The $B_{2g}$ spectrum exhibits a linear low frequency dependence
and a smeared gap. Finally the $A_{1g}$ channel is sensitive to both nodes
and gap because the Raman vertex $\gamma _{{\bf k}}$ doesn't vanish $\ $in
the $k_{x}=\pm k_{y}$, $k_{x}=0$ and $k_{y}=0$ regions. The $A_{1g}$ channel
have to show a linear low $\omega $ behaviour. This remains available when
we take into account the screening.\ 

\ To provide a more realistic account of experimental results with respect
to the theory we have included life-time effects. We have chosen to include
electronic damping by convoluting the bare\ spectra with a Lorentzian
function. Depending on the electronic correlation model, the reduced half
width of the lorentzian can be taken as a constant ($\Gamma /2\Delta
_{0}=a_{0})$, proportional to $\omega $ $\ (\Gamma /2\Delta
_{0}=a_{1}(\omega /2\Delta _{0})$ \ or proportional to $\omega ^{2}$ ($%
\Gamma /2\Delta _{0}=a_{2}(\omega /2\Delta _{0}^{\text{ }}$)$^{2}$).

The $B_{1g}$ experimental spectrum exhibits a nearly cubic frequency
dependence up to $700$ $cm^{-1}$. Such a behaviour imposes, whatever the
type of damping, to use small values for the $a_{0}$, $a_{1}$, $a_{2}$
parameters. For $a_{0}$ $=a_{1}=a_{2}=0.06$ we obtain satisfactory fits for
the $B_{1g}$ spectrum, conversely if we increase by a factor 2 the reduced
parameters, fits are in disagrement with experiments. As already explained 
\cite{10aa} we have chosen to display fits for $\Gamma (\omega )=a_{1}\omega 
$ $.$ The calculated curves of the $B_{1g}$ and $B_{2g}$ spectra for three
values of the $a_{1}$ parameter are shown in Fig. 3. The half width at half
maximum of the $B_{1g}$ peak is seen to double as the $a_{1}$ parameter is
increased by a factor $5$. This dependence reduces the choice of the $a_{1}$
parameter to values close to $0.06$ in order to be in agreement with the
experimental data. This value of \ the damping parameter $a_{1}$ should be
compared with the value of $0.15$ previously obtained on more oxygen
disordered $Hg-1223$ \cite{10aa}. The fact that in this latter case $Tc$ is
increased may appear paradoxical. We are forced to conclude, in that case,
that the effect of doping overwhelms \ the effect of scattering by disorder.
A similar conclusion can be reached from infrared-measurements \cite{13c}.

The theoretical spectra in $B_{1g}$ , $B_{2g}$ and $A_{1g}$ screened
channels for $a_{1}=0.06$ are displayed in Fig. 4. Fits have been calculated
from the strongly UD $Hg-1223$ data. The 2$\Delta _{0}$ gap has been defined
from the $B_{1g}$ spectrum since the $B_{1g}$ vertex probes directly the $%
\Delta _{0}$ amplitude. We have chosen to adjust the lower part of the
calculated curves to the experimental data. Below $300$ $cm^{-1}$ we find a
very good agreement between theoretical and experimental data for all the
three channels and up to the gap for the $B_{1g}$ channel. However the fits
are not satisfactory at higher energy (above $300$ $cm^{-1}$) for the $%
B_{2g} $ and $A_{1g}$ channels. Firstly the $B_{2g}$ maximum of the
calculated spectrum ($700$ $cm^{-1}$) is clearly far away from the $B_{2g}$
experimental one ($450$ $cm^{-1}$) as it is seen in Fig. 4. As a consequence
the calculated frequency ratio between the $B_{1g}$ and $B_{2g}$ peaks
positions (1.2) do not correspond to the experimental one (1.6) obtained
from the $Hg-1223$ spectra. Secondly the calculated curve of the $A_{1g}$
screened spectrum is a broad peak centered around $700$ $cm^{-1}$ whereas we
get experimentally a sharper (width at half maximum : $200$ $cm^{-1}$) and
more intense peak near $550$ $cm^{-1}$.

Concerning the first point ($B_{2g}$ spectrum) we propose to expand the $%
B_{2g}$ vertex to higher order as was already done for the $A_{1g}$ vertex.
The justification of this procedure is based on the fact that the Fermi
surface is anisotropic. The $B_{2g}$ vertex is expected to be more sensitive
to additionnal harmonic terms than the $B_{1g}$ vertex because the $B_{2g}$
vertex probes a region where the gap changes drastically (nodes of the gap).
It is then possible to reconcile experiments to the $d_{x^{2}-y^{2}}$ model
by expanding the $B_{2g}$ vertex to the next order of the Fermi surface
harmonics.\cite{1} Indeed including higher order Fermi surface harmonics in
the Raman vertices change the relative positions of the $B_{1g}$ and $B_{2g}$
peaks. Cardona et al.\cite{17} showed that the $B_{1g}$ peak shifts down
when varying the $\alpha $ parameter related to the contribution of the
second Fermi surface harmonics : $\gamma _{B_{1g}}\left( Cos(2\theta
)-\alpha Cos(6\theta )\right) $. By taking $\ \gamma _{k}^{B_{2g}}=$ $\gamma
_{B_{2g}}\left( Sin(2\theta )-\beta Sin(6\theta )\right) ,$we find that the $%
B_{2g}$ \ peak softens as $\beta $ increases (see Fig. 5). As expected this
softening is even larger when we consider the $B_{2g}$ vertex than for the $%
B_{1g}$ vertex. \ The calculated curve of the $B_{2g}$ spectrum obtained for 
$\beta =0.4$ is displayed on Fig.4. We find an agreement between our
calculated and experimental curves up to $700$ $cm^{-1}$. These results are
consistent with the $d_{x^{2}-y^{2}}$ model under considering the deviation
away from the cylindricity of the Fermi surface when calculating the
vertices. The deviation of the Raman vertex is assumed larger than the one
of the wave-vector on the Fermi surface. $(\delta \gamma )_{F}>>(\delta
k)_{F}$. Best fits could be further improved by taking the electronic
scattering of the normal state but this is not done here.

Concerning the second point ($A_{1g}$ spectrum) the Coulomb screening
present in the Raman scattering process is might be not as efficient as it
has been considered. Strong mass fluctuations induced by inter-layer
coupling or several sheets of $CuO_{2}$ planes as proposed elsewhere,\cite
{17} could decrease the $A_{1g}$ scattering efficiency. On the other hand
expansion to higher order of the $A_{1g}$ vertex have to be explored.

\bigskip

\section{Conclusion}

In conclusion we have specially investigated strongly under-doped $%
HgBa_{2}Ca_{2}Cu_{3}O_{8+\delta }$ single crystals in order to reduce the
oxygen disorder in the $Hg-plane$ taking care to eliminate parasitic phases
on the crystal surface. In these improved experimental conditions we are
able to report with good accuracy the pure electronic Raman scattering
(without subtraction of phonons) in the $B_{2g}$, $B_{1g}$, $A_{1g}+B_{2g}$
and $A_{1g}+B_{1g}$ channels. The $d_{x^{_{^{2}}}-y^{2}}$ model fits quite
well with the low energy part (below $350$\ $cm^{-1}$) of the spectra for
the $B_{2g}$ and $A_{1g}$ symmetries and up to the gap energy for the $%
B_{1g} $ one. However, in order to reconciliate with the $%
d_{x^{_{^{2}}}-y^{2}}$ model, the upper part of the $B_{1g}$ and $B_{2g}$%
spectra (above $350$ $cm^{-1}$), we have to expand the $B_{2g}$ vertex to
the next order of Fermi surface harmonics. The sharp and intense electronic
maximum in mixed $A_{1g}$ symmetries is not easy to interpret in the
framework of the $d_{x^{_{^{2}}}-y^{2}}$ model.\ This last point raises new
questions on the $A_{1g}$ screening efficiency. Experimentally, this work on
near stoichiometric $Hg-1223$ compounds as compared to previous work on less
stoichiometric compounds highlights the importance of oxygen disorder on the
low energy excitations as revealed by the frequency dependence, cubic or
linear of the electronic Raman spectra. This surprisingly strong effect
already reported in a number of high$-Tc$ cuprates is understood in only a
qualitative way and bears directly on the dynamic of the low lying
electronic states in the superconducting state.

\ \ \ \ 

\section{Acknowledgements}

We thank T. Timusk, J. J. McGuire, R. Combescot, N. Bontemps, N. Sandler, B.
Jusserand, M. T. B\'{e}al-Monod and R. Lobo for very fruitful discussions.
We are grateful to M. Jouanne and J. F. Morhange for allowing us to use
Raman spectrometer (LMDH, Paris VI). We also wish to acknowledge S.
Poissonnet and L. Schmirgeld-Mignot for the chemical analysis by electron
probe.

{\bf {Figure Captions}}$\newline
$

FIG. 1 DC magnetization in FC and ZFC of strongly and slightly UD Hg-1223
single crystals. The curve of the slightly UD Hg-1223 is taken from the
reference \cite{10aa}.

FIG. 2\ Imaginary part of the electronic response function of the strongly
UD Hg-1223 single crystal\ in the $B_{2g}$, $B_{1g}$, $A_{1g}+B_{2g}$ and $%
A_{1g}+B_{1g}$ channels. Solid and dashed lines correspond respectively to
the superconducting and the normal state obtained from $T=13$ $K$ and $T=295$
$K$. No phonon subtraction was necessary.

FIG. 3 Theoretical curves of the $B_{1g}$ and $B_{2g}$ spectra for different
values of the $a_{1}$ damping parameter ($\Gamma =a_{1}\omega )$.

FIG. 4 Comparison between the experimental data and the $%
d_{x^{_{^{2}}}-y^{2}}$ model. The calculated curves obtained from the $%
B_{1g} $, $B_{2g}$ and $A_{1g}$ vertices defined in Eq. (3) are shown in
solid lines. The dashed line corresponds to the $B_{2g}$ calculated curve
including the second order expansion of the $B_{2g}$\ vertex ($\beta =0.4)$.
For all the calculations $a_{1}=0.06$.

FIG. 5 Theoretical curves of the $B_{2g}$ response for $\beta =0,$ $0.4,$ $%
0.8;$ ($a_{1}=0.06$).

\bigskip


\begin{references}
\bibitem{9a}  This last observation was made by inserting a video camera
just behind the entrance slit of the spectrometer which allowed us to
precisely choose ($50$ $\mu m$) the laser spot on the crystal surface.
\bibitem{12aa}  The uncertainty inherent to this procedure is believed to be
negligeable.
\bibitem{12bb}  In reference \cite{10aa} we quoted $\delta $ less than 0.1
for $Tc=126$ $K$. This value was later corrected by thermogravimetry
measurements to $\delta =0.16$. Hence the crystal used in this previous work
belongs to slightly UD system.
\bibitem{13aa}  The dark current is constant and its intensity corresponds
approximatively to $30\%$ of the B$_{2g}$ signal at $13$ $K$ and above $800$ 
$cm^{-1}.$
\bibitem{13b}  There is no correction for cross polarization ($B_{1g}$ and $%
B_{2g}$) and small contribution is removed by extrapolation below $50$ $%
cm^{-1}$for parallel polarizations ($A_{1g}$).
\bibitem{14c}  We have estimated the Raman intensity as a function of the $%
BaO_{2}$ phase percent. It was done from the Raman spectrum of the $BaO_{2}$
powder (in the bulk form) obtained in the same experimental conditions as
the $B_{2g}$ and $B_{1g}$ spectra in the S state. $100\%$ of illuminated
bulk of $BaO_{2}$ gives $500$ arb. units at $300$ $cm^{-1}$.
\end{references}
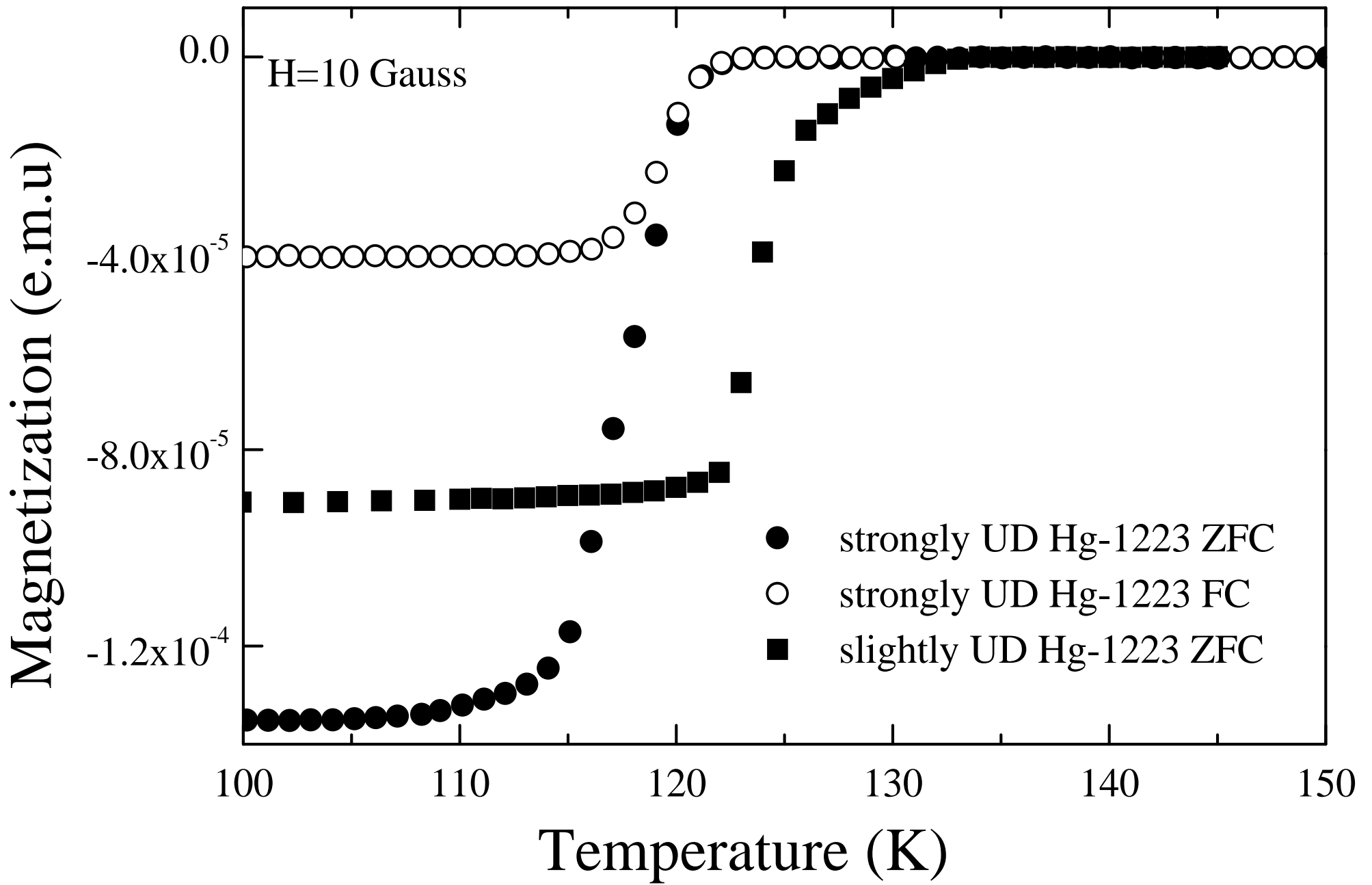
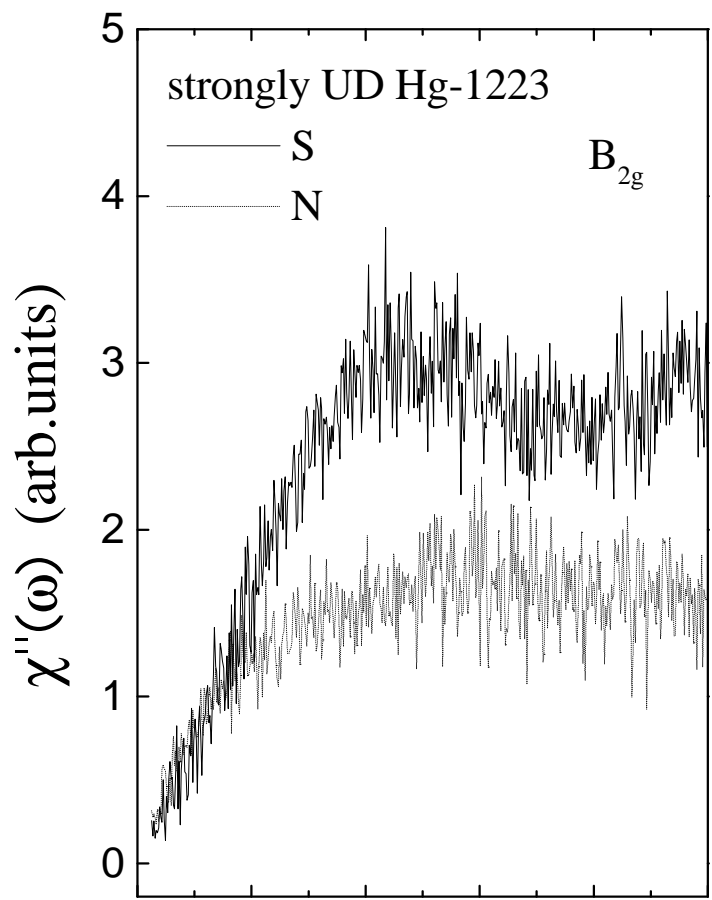
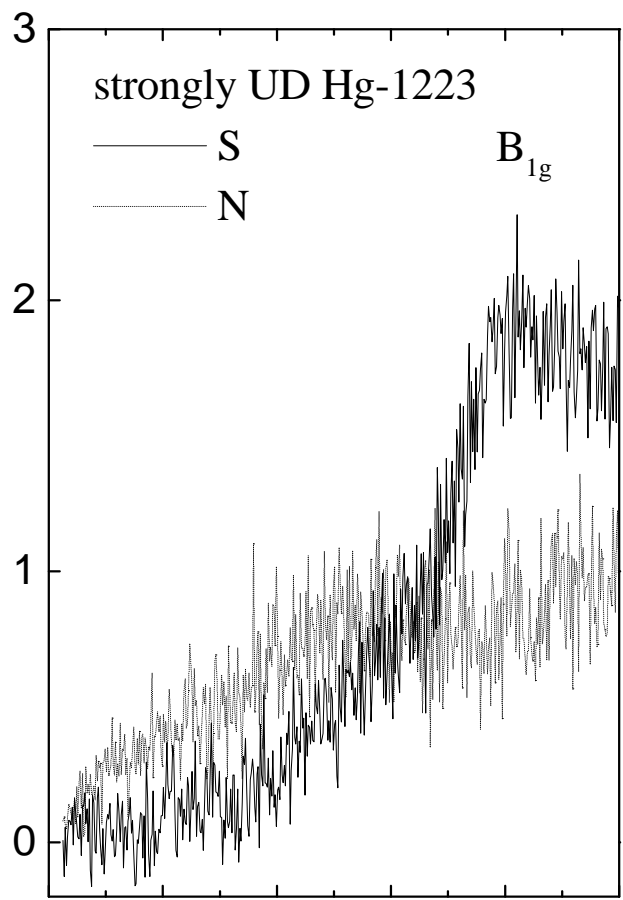
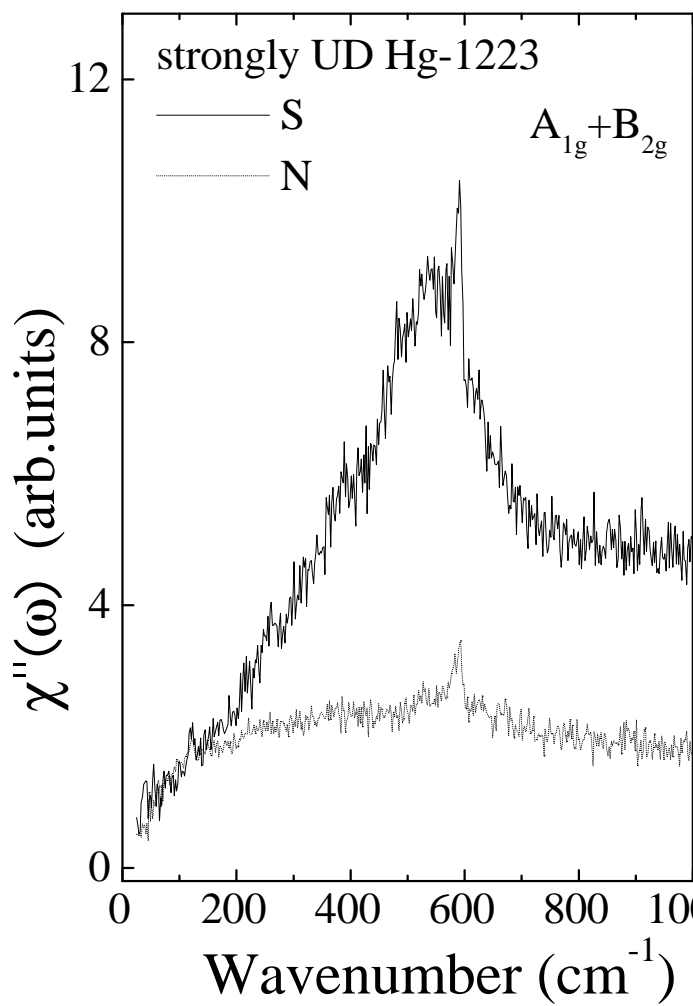
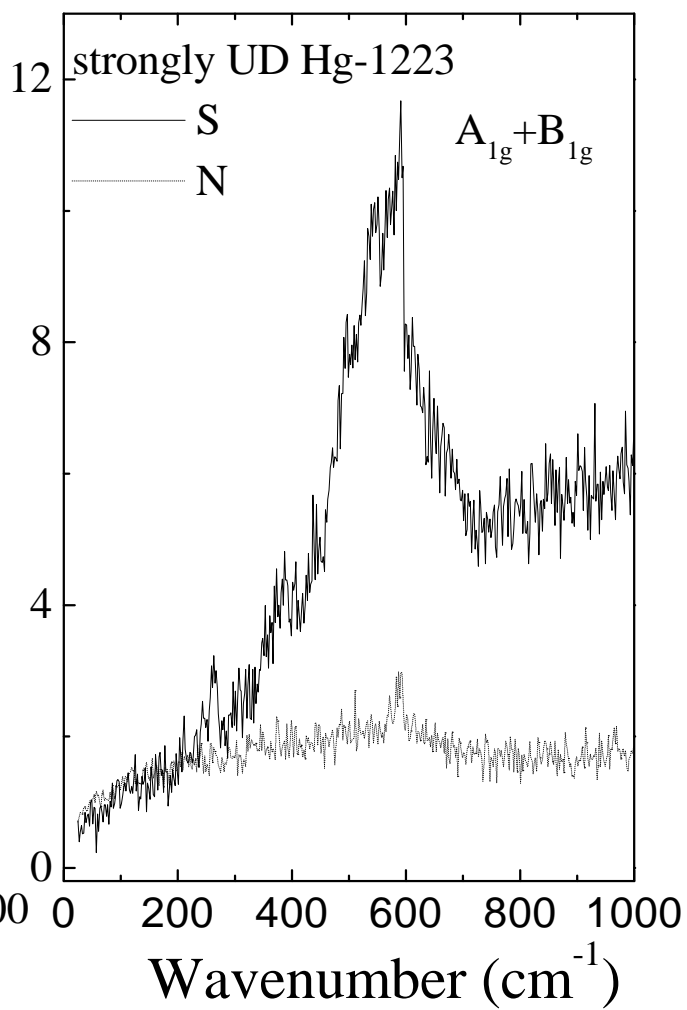
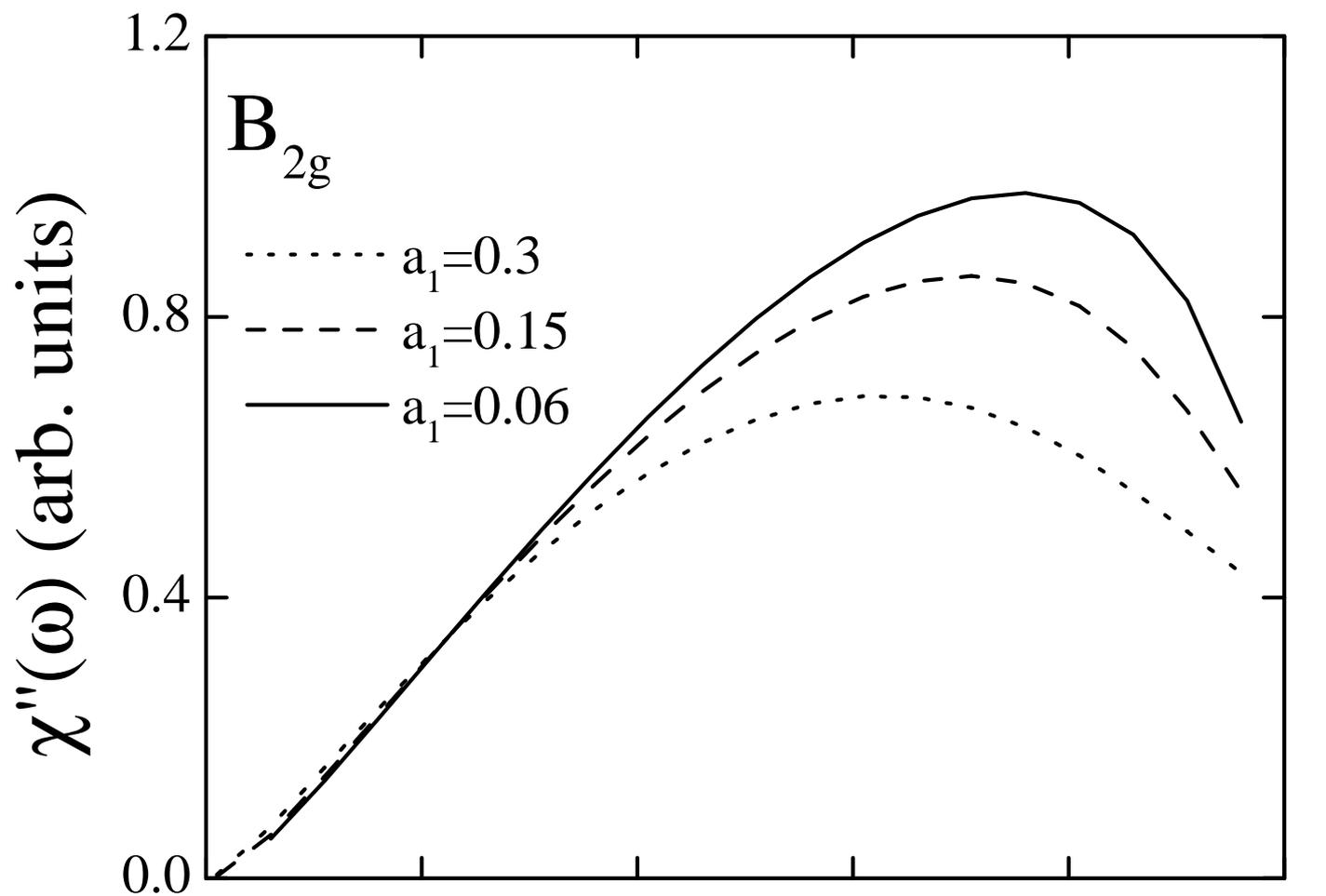
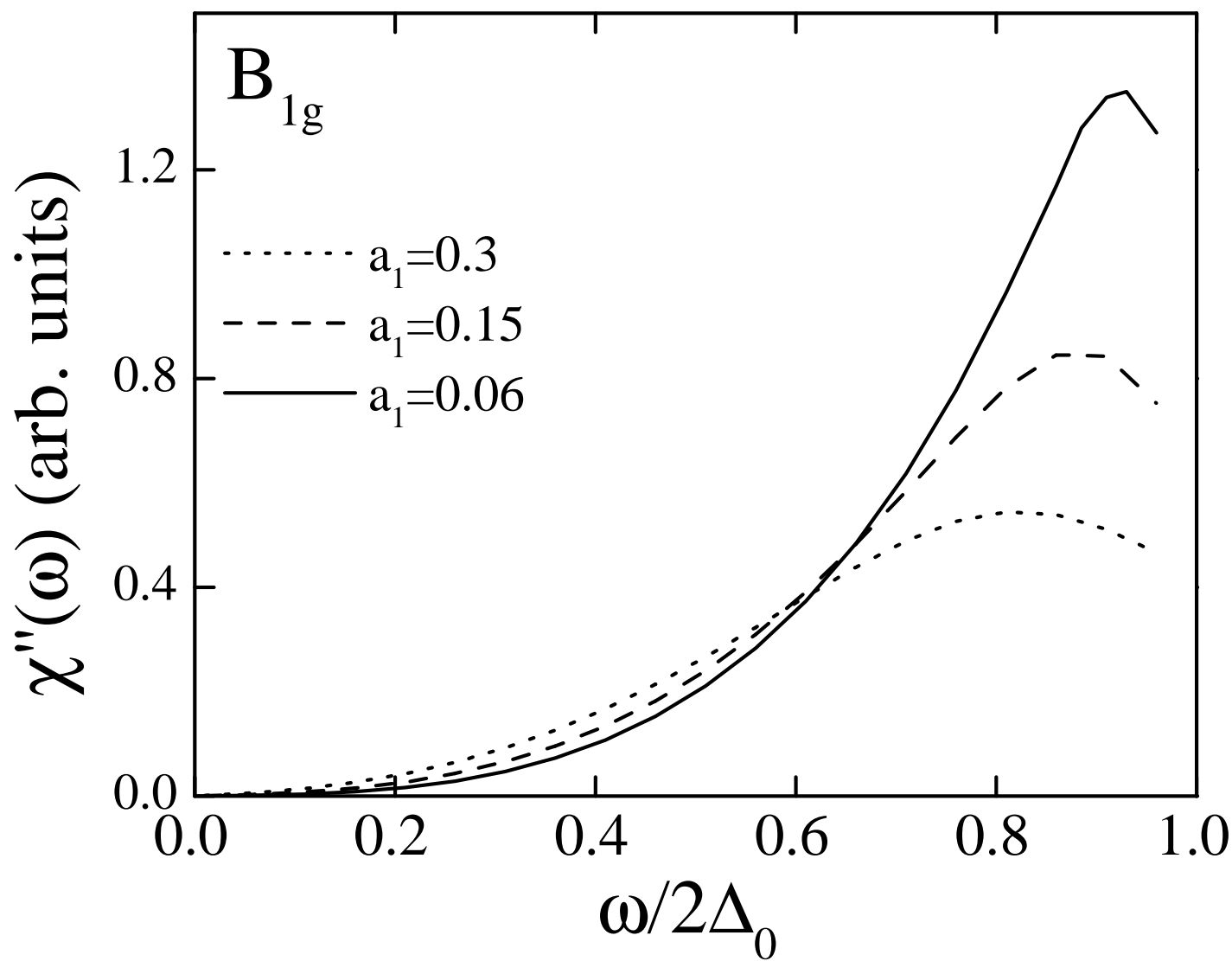
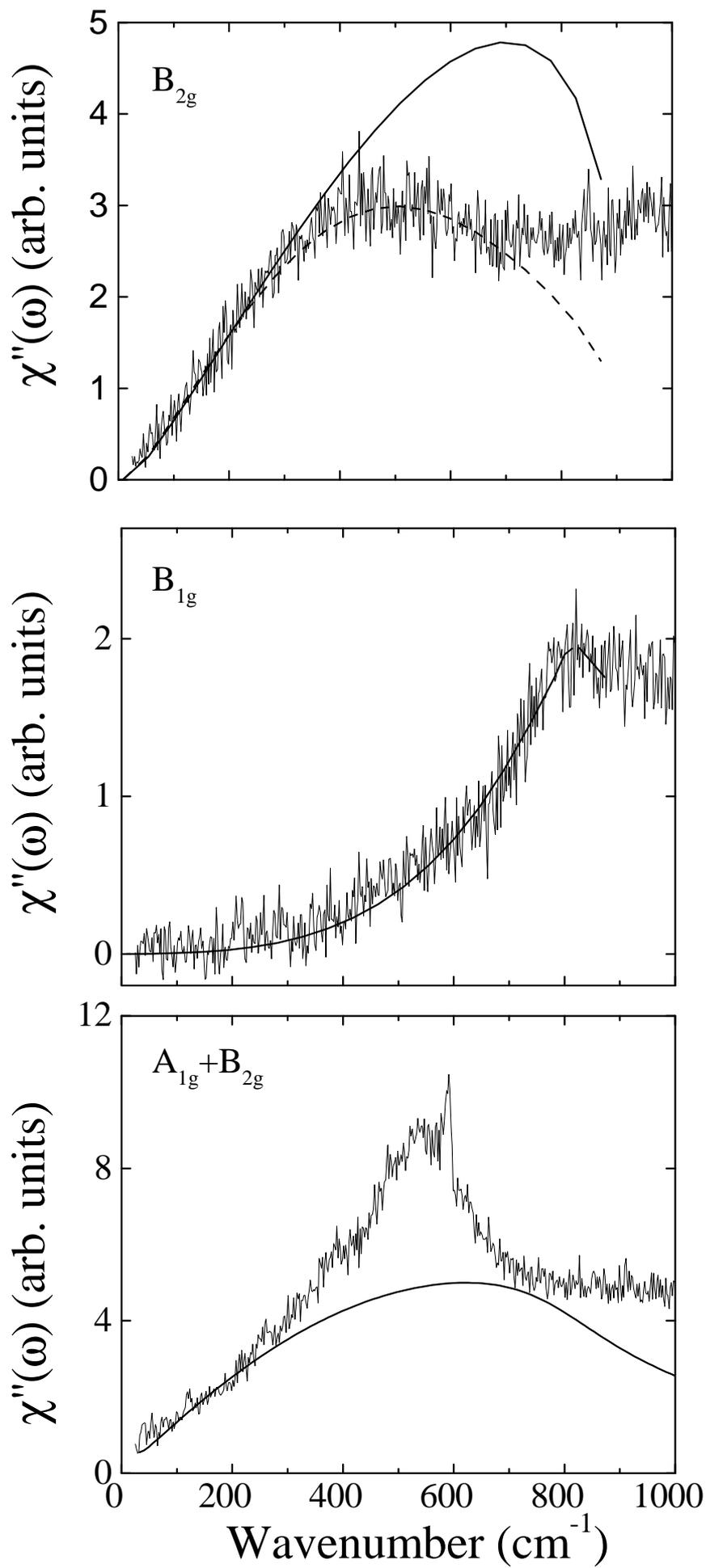
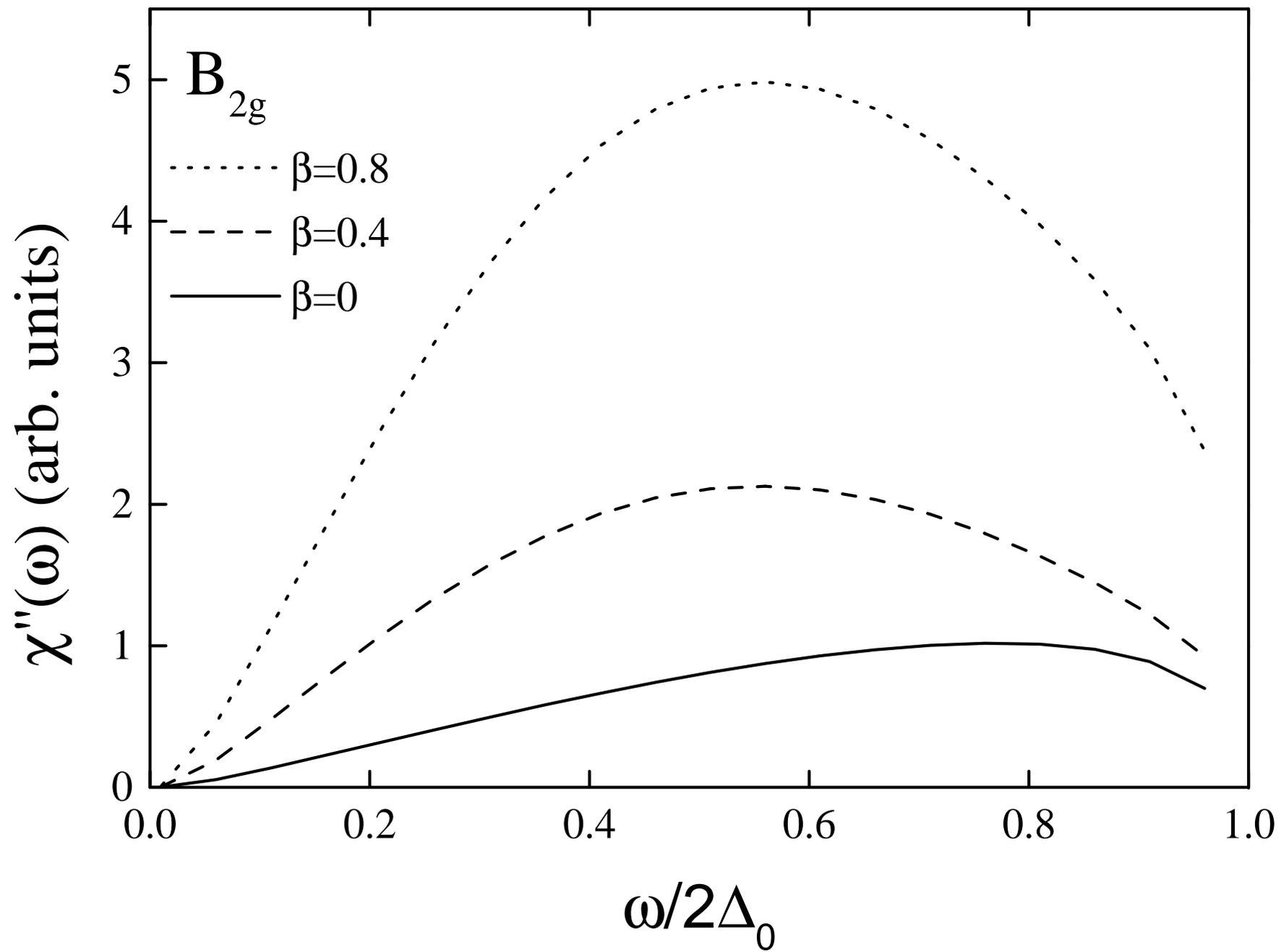
\end{document}